# Dominant Sets Based Band Selection in Hyperspectral Imagery

Onur Haliloğlu, Ufuk Sakarya, B. Uğur Töreyin, Orhan Gazi

onur.haliloglu@msspektral.com, usakarya@yildiz.edu.tr, toreyin@itu.edu.tr,
orhan.gazi@ankaramedipol.edu.tr

*Abstract*— **Hyperspectral imagery is composed of huge amount of data which creates significant transmission latencies for communication systems. It is vital to decrease the huge data size before transmitting the Hyperspectral imagery. Besides, large data size leads to processing problems, especially in practical applications. Moreover, due to the lack of sufficient training samples, Hughes phenomena occur with huge amount of data. Feature selection can be used in order to get rid of huge data problems. In this paper, a band selection framework is introduced to reduce the data size and to find out the most proper spectral bands for a specific application. The method is based on finding "dominant sets" in hyperspectral data, so that spectral bands are clustered. From each cluster, the band that reflects the cluster behavior the most is selected to form the most valuable band set in the spectra for a specific application. The proposed feature selection method has low computational complexity since it performs on a small size of data when realizing the feature selection. The aim of the study is to find out a general framework that can define required bands for classification without requiring to perform on the whole data set. Results on Pavia and Salinas datasets show that the proposed framework performs better than the state-of-the-art feature selection methods in terms of classification accuracy.**

## I. Introduction

With the development of hyperspectral imaging that provides hundreds of narrow contiguous bands, more information can be gathered in the detection of materials and objects as compared to multispectral imaging. However, hyperspectral imagery has some drawbacks such as huge data volume and redundancy. Huge data volume requires large processing time, bigger data storage capacity, and higher transmission rates. It is not always possible to provide the requirements of huge data volume. Therefore, dimension reduction techniques are applied to reduce the data size.

Dimension reduction techniques can be divided into two main groups: feature extraction and feature selection. In feature extraction methods, input data is transformed to a different space where a new set of features are innovated in order to increase the discriminative ability. On the other hand, in feature selection techniques, a subset of the original input data is selected based on a set of criteria. Unlike feature extraction, feature selection does not alter data during data transformation.

To avoid the ill-conditioned formulas arising due to the high-dimensional data space of hyperspectral images (HSIs) dimensionality reduction using ensemble distance metric learning is proposed in [1]. In ensemble distance metric learning, by reducing the dimensionality of the data, discriminative information use is enhanced. In [2], a spectral-spatial feature extraction method based on ensemble empirical mode decomposition for Hyperspectral image classification is proposed. A feature extraction method using multidimensional spectral regression whitening is proposed in [3], and in this method, the data passes through a number of steps involving, entropy rate segmentation, spectral regression whitening, classification, and voting. Iterative ReliefF (I-ReliefF) is a single domain-based feature selection method to determine a feature subset used for classification. Single domain-based feature selection approaches suck for huge data sizes. In this case, methods based on more domains are utilized. Cross-scene feature selection algorithms employ two domains, and it is useful for large data sizes, however, this algorithm suffers from spectral shift problems.

In [4], the authors merge I-ReliefF and cross-scene algorithms and introduce a cross-domain I-ReliefF algorithm which shows better performance than that of the I-ReliefF and cross-scene algorithms. In [5], making use of the spatial information, the authors proposed patch-based and tensor patch-based, approaches to be utilized by graph-based dimensionality reduction methods. The proposed methods are called weight-modified tensor locality preserving projections and weight-modified tensor neighborhood preserving. It is reported in [5] that the proposed approaches outperform the up-to-date algorithms. In hyperspectral imaging, spectral and spatial information can be concatenated to obtain longer feature vectors and in sequel dimensionality reduction can be performed, and finally, classification can be achieved. However, such a rough approach to obtain the longer feature vectors faces many weaknesses due to the inappropriate use of the different data vectors in a longer vector. To alleviate the weak points of vector concatenation, in [6] authors proposed a more smart approach to benefit from different types of feature vectors in an intellectual manner. In [6], authors transform spectral-spatial feature to another feature space and the most significant original features of data vector are used for classification.

Band selection methods are the most popular feature selection methods that select the most valuable bands and remove the redundant bands based on a set of criteria [7, 8, 9].

The hyperspectral band selection technique can be ranking [10, 11] or clustering-based [12, 13, 14]. Ranking-based methods are utilized for detecting the most informative and distinctive bands. However, there is a drawback that ranking-based methods can select redundant bands since correlation among selected bands is not considered in this method. On the contrary, clustering-based methods take correlation among the bands into account. These methods firstly cluster bands based on determined criteria and then select one band from each cluster as a representative one.

Graph-based representations have become the state of the art approaches especially for solving clustering problems in pattern recognition. There are several advantages in using graphs instead of feature vectors for object representation [15]. Shi and Malik introduced normalized cuts method, which was based on a generalized eigenvalue problem to obtain graph partitions according to normalized cuts criterion [16]. On the other hand, Pavan and Pelillo proposed a dominant sets method (DSM) in which a graph-theoretic definition of a cluster was introduced [17]. The DSM is closely related to the maximum clique problem in graphs as well. Pavan and Pelillo [17] demonstrated the relation between DSM and maximum clique when weighted undirected graph is induced to un-weighted undirected graph. In the literature, the dominant set technique is utilized to solve distinct clustering problems [18, 19]. The main idea of these methods is to obtain a graph-based representation of a pattern recognition problem by mapping the graph elements with vertices and mapping the relationship of the graph elements with weighted edges. Therefore, the problem is reduced to solving a graph-based optimization problem. In this paper, we propose novel graph-based representations to be used for hyperspectral band selection. To the best of the authors' knowledge, it is the first study to utilize DSM for hyperspectral image analysis.

Hyperspectral image classification is an important application in agriculture, forestry, geology, ecological monitoring, and disaster monitoring [20]. According to the specific application area, materials or elements under consideration may vary. A set of selected bands may be useful in one application domain; on the other hand, the same set may not provide any use in another application domain. In a typical real life scenario, a few numbers of distinct classes are required to be discriminated for a particular application using only a limited number of training data. The main challenge is to detect the predetermined classes among a larger number of distinct classes. To achieve this goal, a framework is required to obtain an optimized set of bands that has not only inter-class discriminative information but also intra-class coherence for a specific application. DSM has the advantage of graph-based representation simplification and aims to increase intra-cluster heterogeneity while decreasing interclass heterogeneity. Therefore, it is a good idea to combine DSM approach with band selection in order to handle the main challenge in the real life scenario. In this paper, we propose a novel framework based on dominant sets approach i.e., dominant sets based feature selection method (DSbFSM) in order to overcome the above mentioned problem.

## II. THE PROPOSED APPROACH

DSbFSM is a graph-based clustering band selection method. It aims to cluster the bands by dominant sets approach and selects the most valuable band from each cluster. In clustering band selection methods, clustering can be realized considering different criteria. In [12] mutual information is used to cluster the spectral bands whereas the mean average deviation criterion is employed in [13] for clustering the bands. Rather than finding the most informative bands among the whole spectra, it is crucial to find the most distinctive bands for predetermined classes. Therefore, DSbFSM computes the discrimination performances of the bands for determined classes in a specific application in order to utilize the result as a band clustering criterion. To evaluate the discrimination performance of each band, it is meaningful to use a clustering method. Therefore, the classification performance of each band is again evaluated by dominant sets approach.

The DSbFSM has two stages. In the first stage, the pixels in each band are clustered into groups. These clusters are matched with known classes by using the ground truth and the clustering performance of each band is evaluated for each class. Therefore, the first stage puts forward to the performance of each band for each class. The second stage uses the results of the first stage in order to cluster the bands of hyperspectral image. The most representative band in each cluster is selected. As a result, a set of bands that contains the most discriminative information according to training data and has the least redundancy is selected. Shortly, the first stage is the performance evaluation of bands with respect to defined classes whereas the second stage is the band selection process using the outputs of the first stage. In both stages, dominant sets based clustering approach is performed. In the former, DSM is utilized to cluster the pixels. On the other hand, in the latter DSM is utilized to cluster the bands. The DSbFSM algorithm is explained in **Error! Reference source not found.**.

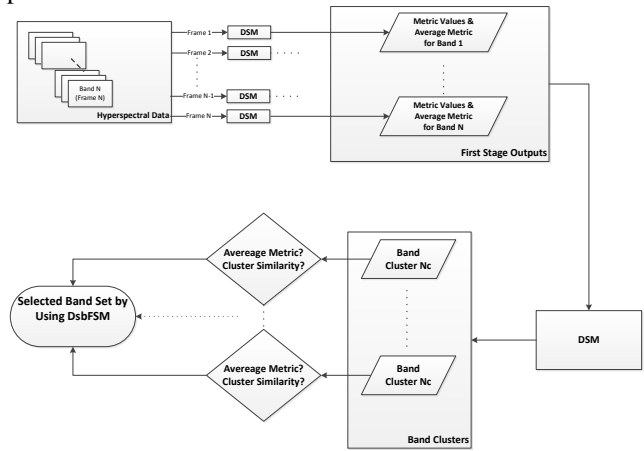

Fig. 1 DSbFSM Algorithm.

The first and second stages of the proposed method will be explained in detail in the following sections. The main steps of the proposed method are:

1. DSM is applied at each frame that belongs to a specific band for clustering.

2. Metric values of each frame are evaluated for each class, and, a metric associated with each frame is evaluated at each class and average metrics for all the classes are calculated. The average metric is the mean value of the metrics for each class.

3. The calculated metrics of frames are utilized for finding out the similarity among frames in order to cluster the bands of the hyperspectral image using the DSM.

4. The band that represents its cluster properties **the** most i.e., the band that has high interclass similarity and high average metric value, is selected from each band cluster. The selected bands form the set of bands determined by the proposed feature selection method.

Both stages of the proposed scheme are based on the dominant sets method. Fundamentals of the dominant sets method are provided in the sequel.

## III. FUNDAMENTALS OF DOMINANT SETS METHOD

Clustering problems occur in many areas of engineering such as image segmentation, computer vision, data mining etc. Pavan and Pelillo [17] proposed a graph theoretic approach combining the concepts of cluster and dominant set of vertices. This dominant sets method aims to increase the inter-cluster homogeneity while decreasing the intra-cluster homogeneity.

In this framework, the first step is to find out the similarities between the graph vertex pairs $(i,j)$ where $1 \leq i,j \leq N$, and $N$ is the total number of vertices in the graph. The similarity measure between the vertices $i$ and $j$ are indicated by $W(i,j)$ such that $0 \leq W(i,j) \leq 1$. The similarity measures, $W(i,j)$, between all vertex pairs form the similarity matrix $W$. The similarity matrix $W$ contains information about the connectivity between the vertex pairs; the vertex pair $(i,j)$ are said to be connected if $W(i,j) > 0$, otherwise $(i,j)$ pair is not connected. A graph $G(V,E,W)$ is constructed by a set of vertices $V = \{1, \dots, N\}$, a set of edges $E = \{(i,j)|W(i,j) > 0\}$, and the similarity matrix $W$, with the assumption that there are no self-loops in the graph.

Let $S \subseteq V$ be a nonempty subset of vertices and $i \in S$, where $1 \leq i \leq |S|$ and $|S|$ is the total number of vertices in $S$. The average weighted degree of $i$ with respect to $S$, $M_S(i)$, is defined as:

$$M_S(i) = \frac{1}{|S|} \sum_{j \in S} W(i,j) \qquad (1)$$

$M_S(i)$ is a measure of the similarity of the vertex $i$ with respect to other vertices in $S$. A similar function, $\varphi_S$, is defined to measure the similarity between vertices $j \notin S$ and $i \in S$, with respect to the average similarity between vertex $i$ and its neighbors in $S$.

$$\varphi_S(i,j) = W(i,j) - M_S(i) \qquad (2)$$

The total weight of subset $S$ is defined as:

$$A(S) = \sum_{i \in S} a_S(i) \qquad (3)$$

where $a_S(i)$ is the weight of $i \in S$, and it is recursively evaluated as:

$$a_S(i) = \begin{cases} 1, if |S| = 1, \\ \sum_{j \in S \setminus \{i\}} \varphi_{S \setminus \{i\}}(j,i) a_{S \setminus \{i\}}(j), otherwise. \end{cases} \qquad (4)$$

which is the overall similarity amount between vertex $i$ and the vertices of $S \setminus \{i\}$ with respect to the overall similarity among the vertices in $S \setminus \{i\}$. A non-empty subset of vertices $S \subseteq V$ such that $A(U) > 0$ for any non-empty $U \subseteq S$, is said to be dominant, if:

1. $a_S(i) > 0$, for all $i \in S$,
2. $a_{S \cup \{i\}}(i) < 0$, for all $i \notin S$.

The resulting two statements mentioned above correspond to the cluster conditions that are related to inter-cluster homogeneity and intra-cluster inhomogeneity respectively. The dominant sets determined satisfying the two criteria can be classified as clusters. In practice, direct implementation of this framework is not feasible because of its computational complexity. To solve this problem, a continuous optimization technique, known as *replicator dynamics* can be used [17]. This technique starts from an arbitrary initial state and finally reaches to a maximum value of the replicator equation with some iteration. In order to find out the dominant set of a weighted characteristic vector of the vertices $x$, the iterative equation in (5) is used [17]:

$$x_i(t+1) = x_i(t) \frac{(Wx)_i}{x(t)^T W x(t)} \qquad (5)$$

In (5), $t$ is the iteration index, and $T$ is the transpose operator. The iteration is terminated if the difference between two consecutive iterations of $x$ is below the predetermined threshold. Once the dominant set is detected, the remaining vertices can be subjected to the same equation in order to find new dominant sets i.e., clusters. In this way, clustering can be realized for the data.

## IV. DOMINANT SETS-BASED FEATURE SELECTION METHOD (DSbFSM)

This feature selection method proposes a selected set of bands for discriminating the classes under concern. In order to achieve this, a small part of the hyperspectral image that contains sufficient samples for the interested classes is selected. The DSbFSM will be performed on the selected part of the whole image. Therefore, it is aimed to find out a general framework that can define required bands for classification without performing on the whole data set. Indeed, the selected set of bands can be used to discriminate the classes under concern even for another region that contains not only the same classes but also contains other distinct classes.

*A. First Stage in the Proposed Method: Evaluating Band Performances*

In the proposed approach, the first step is to evaluate the performance of bands with respect to defined classes based on a particular metric. In this study, F-measure metric [15] is used to quantify the classification performance of bands. In order to evaluate the discrimination performance of bands, pixels in each frame that belongs to a specific band are clustered by DSM. Then, the clustered pixels are classified by the ground truth. The cluster is appointed to the class that has the highest number of member pixels in the cluster. As a result, each band frame is classified. Classification performance is evaluated using F-measure metric [15].

Let $H_f$ be the $L \times M \times N$ hyperspectral image i.e., the hyperspectral cube has $L$ bands, $M$ pixels in the image frame row, and $N$ pixels in image frame column. Let $H$ be the partition of $H_f$ including sufficient samples of the classes under concern. That is, $H$ is also a hyperspectral image having $L$ bands and $K$ number of pixels such that $K < M \times N$. Therefore, $H$ can be expressed as:

$$H = \{H_l\}_{l=1}^{L}$$

where $H_l$ is the $l^{th}$ band of the hyperspectral image and it is a $K$ pixel image frame.

In this stage, each frame that belongs to a specific band i.e., $H_l$, is subjected to DSM to form the clusters. The first step is to form the similarity matrices for each band. Let $W_l$ be the $K \times K$ similarity matrix for the $l^{th}$ band and $Wl(i,j)$ is the element of the similarity matrix that represents the similarity between the $i^{th}$ and $j^{th}$ pixels, $1 \leq i,j \leq K$.

In order to form the similarity matrix, it is required to have a similarity measurement function. It is determined that this function is related to both spatial and spectral information. Therefore, $W_l(i,j)$ is defined as:

$$W_l(i,j) = e^{-C_d d_{ij}^2} e^{-C_s |r_l(i) - r_l(j)|} \quad (6)$$

where $C_d$ and $C_s$ are spatial and spectral information parameters respectively. Once these parameters are determined they are used for every $W_l(i,j)$. $r_l(i)$ is the reflectance value of $i^{th}$ pixel for the $l^{th}$ band. Indeed, $r_l(i)$ is an element in $H_l$. The spatial distance in the image frame between pixels $i$ and pixel $j$ is represented as $d_{ij}$.

After forming the similarity matrix, the dominant set is iteratively calculated using (5). The threshold $Th$ defines the number of iterations. The remaining part of the first stage of the proposed method is given in the following steps:

*Step-1*. The initial value of $x_l$ is a $K$ size array consisting of all ones. It means that initially all pixels of the $l^{th}$ band is in the dominant set.

*Step-2*. Calculate the dominant set by using (5).

*Step-3*. If $|x_l(t+1) - x_l(t)| > Th$, then $x_l(t) = x_l(t+1)$ and go to *Step-2*. Otherwise, go to *Step-4*.

*Step-4*. The dominant set is finalized. The dominant set elements constitute the cluster. This cluster is assigned to the proper class by using the ground truth. The maximum number of cluster pixels that are affiliated to the same class defines the class of the cluster. Therefore, the pixels in the dominant set are classified.

*Step-5*. The classified pixels are extracted from the image and the similarity matrix elements related to these pixels are eliminated from the similarity matrix $W_l$.

*Step-6*. If there are unclassified pixels, then form the array $x_l$ with a size equal to the number of unclassified pixels consisting of all ones and go to *Step-2*. Otherwise, go to *Step-7*.

After performing the iterative algorithm, all the pixels are classified. The next step is to evaluate the performance of bands for each class with a metric. For the evaluation process, it is important to succeed in both clustering the pixels that belong to the same class in the same cluster and preventing from the mismatch of the cluster pixels with incorrect classes. Therefore, $F$-measure metric [21] is selected to evaluate the performances since it takes not only the detection but also the false alarm into account. It measures the quality of the detected clusters and its value ranges from 0 to 1. $F = 1$ indicates a perfect result.

$F$ is defined as

$$F = \frac{1}{Z} \sum_{C_i \in GT} |C_i| \max_{C_j \in DT} \{f(C_i, C_j)\} \quad (7)$$

where

$$f(C_i, C_j) = \frac{2 \times \text{Re}(C_i, C_j) \times \text{Pr}(C_i, C_j)}{\text{Re}(C_i, C_j) + \text{Pr}(C_i, C_j)}, \quad (8)$$

$GT$ is the ground truth and $DT$ is the dominant set, and $Z$ is calculated as

$$Z = \sum_{C_i \in GT} |C_i|. \quad (9)$$

*Recall* ($Re$) and *precision* ($Pr$) functions are defined as:

$$\text{Re}(C_i, C_j) = \frac{|C_i \cap C_j|}{|C_i|} \quad (10)$$

$$\text{Pr}(C_i, C_j) = \frac{|C_i \cap C_j|}{|C_j|} \quad (11)$$

*Step-7*. As described in *Step-4*, $F$ metric is calculated after the clusters are assigned to the proper class. Calculate the $F$ metric of the $l^{th}$ band for each class by using the equation

$$F_l(i) = \frac{1}{Z} \sum_{C_i \in GT, C_j \in DT} |C_i| \times f(C_i, C_j)$$

for $1 \leq i \leq P$ where $P$ is the number of classes in the ground truth.

*Step-8*. $F_l$ is $l^{th}$ band performance for each class and each element of this array $F_l(i)$ refers to a distinct class in the ground truth. In addition to this, an overall metric is calculated for the $l^{th}$ band denoted by $F_{l,tot}$ which is the mean value of $F_l$.

Steps 1 to 8 are applied to each band of the hyperspectral cube. Therefore, the performances of the hyperspectral bands are evaluated separately.

## B. Second Stage of the Proposed Method: Clustering and Selection of Bands

In the first stage of the proposed method, the performance evaluation of bands with respect to defined classes is realized. In the second stage, the set of bands that mostly discriminates the given classes by using the information coming from the first stage are determined.

Since the output information of the first stage is the input of the second stage, we have to start with $F_l$ for $1 \leq l \leq L$. As mentioned before $F_l$ is the $F$-metric values of the $l^{th}$ band for each class. In fact, $F_{l,tot}$ that is defined as the overall performance of the $l^{th}$ band is also calculated for each band in the stage-1. However, it is not meaningful to select the bands having largest $F_{l,tot}$. Due to the fact that the band which has low overall performance can possess high performance for a distinct band, it is better to use $F_l$ that includes distinct performances for each class.

The main aim of the band selection is to eliminate redundancy. Therefore, it is required to know the redundant bands. Band clustering is a good solution to define the redundant bands. $F_l$ is a signal having $P$ samples and dominant sets technique is applied among all $F_l$, $1 \leq l \leq L$. To apply dominant set technique, similarity measurement matrix $W_B$ is formed. $W_B$ is a $L \times L$ symmetric matix and its elements are calculated as:

$$W_B(i,j) = e^{-C_B(1-corr(F_i,F_j))} e^{-C_V \frac{1}{P} sum(abs(F_i,F_j))} \quad (12)$$

where $C_B$ is the band correlation parameter and $corr(f,g)$ is the correlation of signals $f$ and $g$. Additionally, $C_V$ is the mean square distance parameter and $abs(.)$ is the absolute value function. At first glance, it seems that correlation function is sufficient to form the similarity matrix. However, the distance between band pairs is also crucial to distinguish the bands. The first exponential term of the similarity matrix is related with the correlation between band pairs whereas the second exponential term is related to the mean square distance term in the literature [22] that defines the distance between the band pairs. Besides the correlation function, the distance between band pairs is also crucial to distinguish the bands. Let $A$ be an $R$ dimensional signal and $B$ is another signal that is equal to $c$ times $A$ where $c$ is a constant between 0 and 1. $corr(A,A)$ and $corr(A,B)$ are both equal to 1 unless $B$ is different from $A$. In order to differentiate between $F_l$, mean square distance term shall be used while calculating the similarity matrix.

By using (5), the dominant sets of bands are calculated iteratively. The iterations finalize provided that the distance between two successive iterations is smaller than the threshold $T_B$. When the iterations finalize, a dominant set is found. This dominant set refers to a band cluster consisting of similar and/or redundant bands. The remaining bands are subjected to the dominant sets algorithm described above until no unclustered band exists.

At this stage, there are $N_C$ numbers of band clusters. Therefore, $N_C$ bands have to be selected from each cluster to determine the set of bands formed by the proposed feature selection method. The selected band within the cluster has to satisfy the criteria:
1. The selected band has high $F_{l,tot}$

2. The selected band has highly correlated with other bands within the cluster.

To satisfy the two criteria, band selection parameter is defined as:

$$\eta_{l,z} = F_{l,tot} \times \sum_{j \in D_z} W_B(l,j) \quad (13)$$

where $D_z$ is the band cluster for $1 \leq z \leq N_C$ and $\eta_{l,z}$ is the band selection parameter of the $l^{th}$ band which is located in the $z^{th}$ cluster. The band which has the maximum $\eta_{l,z}$ value is selected from its cluster. The selected bands form the output of the proposed feature selection method.

## V. DATA SET AND PERFORMANCE EVALUATION

### A. Data Set and Ground-truths

Salinas and Pavia scenes [23] are employed in our work. As described earlier, DSbFSM is performed on a small partition of the whole hyperspectral image that contains sufficient samples for the interested classes.

Salinas A data includes six classes. DSbFSM is performed to find out valuable hyperspectral bands for the six classes. Salinas A data is the subset of the Salinas data. Salinas A is approximately 5% of Salinas data when removing the unclassified regions. Salinas data includes 10 more classes but these class samples are signed as unclassified since they are not under concern. The ground truth of Salinas A is shown in Fig. 2.

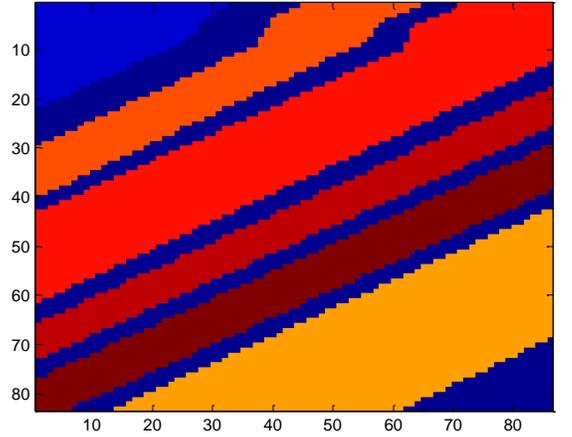

Fig.2 Salinas A ground truth including 6 classes and unclassified regions.

A partition includes trees, bitumen, tiles, meadows, bare soil, and water classes is selected from the Pavia data. This partition is approximately 0.3% of the full Pavia data when removing the unclassified regions. Selected part of Pavia including four classes is shown in Fig. 3.

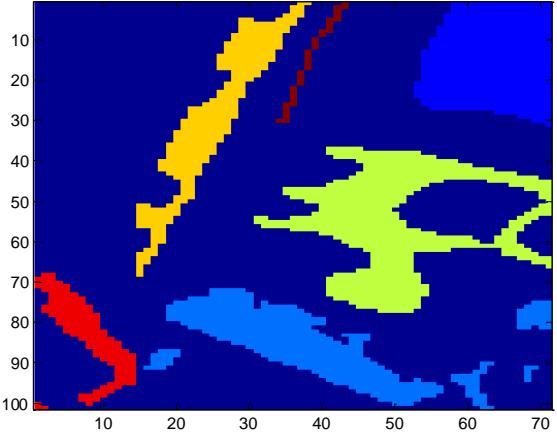

Fig.3 Ground truth of Pavia selected partition including 6 classes and unclassified regions.

## B. Performance Evaluation

DSbFSM is performed on the three distinct hyperspectral data as mentioned in Section V-A. While implementing the DSbFSM, there are four parameters to determine. Two of them are for the first stage $C_d$ and $C_s$. These parameters are found iteratively by using the red edge band in Pavia data as 0.3 and 0.01 for the spatial and spectral parameters respectively. For Salinas data, spatial parameter remains constant and spectral parameter is multiplied by the ratio of the radiance mean values between the performed data and the Pavia data. The remaining two parameters $C_B$ and $C_V$ are for the second stage and taken as 200 and 30 for all hyperspectral data.

While performing the DSbFSM, only the samples of the classes under concern are taken into account. Once DSbFSM is performed on small partition of the whole hyperspectral data and valuable bands are selected, a new hyperspectral image that includes all the samples but contains only the selected bands is formed. SVM is applied to this new hyperspectral data in order to find out the performance of DSbFSM in terms of overall classification accuracy for the classes under concern. In order to evaluate the performance of DSbFSM, Correlation Band Selection (CBS), and Sequential Forward Selection (SFS) [24] band selection methods are implemented. SFS is a suboptimum band selection method that is used in many studies recently. CBS is a simple method that realizes the band selection by considering the correlation among bands. Most uncorrelated bands are selected by CBS. For fair evaluation, the number of selected bands of CBS is equalized to the number of selected bands of DSbFSM. On the other hand, since SFS is an iterative suboptimum method the selected bands of SFS are accepted as they are. Classification Accuracy [25] vs Number of Training Set performances are analyzed for three band selection methods. Number of training set value defines the number of training samples selected for each class. SVM is performed for different number of training sets, and classification accuracy is calculated for 20 times for the same training set number to get the average value. Training set for SVM is randomly selected and includes samples from each class. While comparing the performances of the three band selection methods with SVM, training set for SVM is randomly selected, and the same training set is used for all band selection methods.

## VI. RESULTS

DSbFSM, CBS, and SFS performances for the Salinas data are shown in Fig. 4. It is observed that DSbFSM performance is slightly larger than that of SFS performance. CBS performance is lower than the others especially for small numbers of training sets.

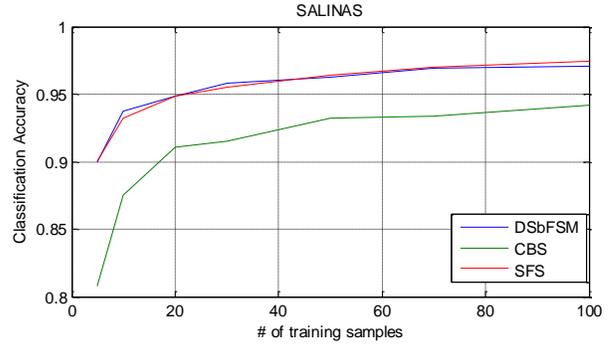

Fig.4 Classification Accuracy vs Number of SVM training set samples performances of DSbFSM, CBS, and SFS for Salinas Data.

DSbFSM, CBS, and SFS performances for the Pavia data are shown in Fig. 5. It is observed that DSbFSM performance is better than SFS and CBS performances. The size of the partition the Pavia data used to realize band selection is really small compared to whole Pavia data. It is investigated that DSbFSM performs well even when the size of the training data for band selection is very low.

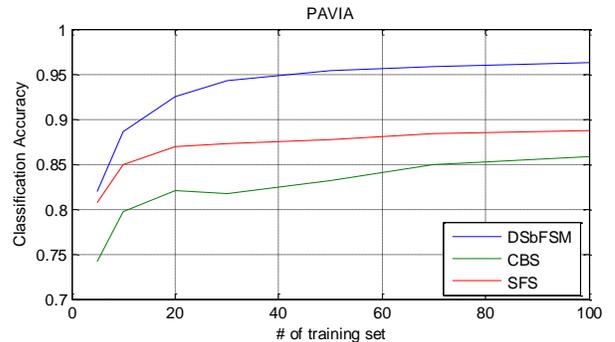

Fig.5 Classification Accuracy vs Number of SVM training set samples performances of DSbFSM, CBS, and SFS for Pavia Data.

DSbFSM is proposed for a general set of selected bands for the classes under concern with high classification accuracies. Moreover, CBS and SFS is time consuming since they perform iterative steps to find out the selected bands. In addition to this, the classification method SVM is run for each iteration when SFS is performed.

## VII. CONCLUSIONS

In this paper, we propose novel methods to decrease the huge data size of *Hyperspectral imagery* which creates significant latency for communication systems.8u The proposed feature selection method DSbFSM performs on a small size of the entire data that includes samples from classes under concern. The performance evaluation of DSbFSM indicates that the proposed set of selected bands by DSbFSM can be used to discriminate the classes under concern even for another region that contains not only the same classes but also contains other distinct classes. Therefore, DSbFSM is a general framework that can define required bands for classification. When compared to the other feature selection methods mentioned in this study such as CBS and SFS, DSbFSM has higher classification performance. Moreover, DSbFSM is time efficient since it performs on small size of data and it does not require to calculate classification accuracy iteratively in order to realize feature selection.